\documentclass[aps,prl,showpacs,twocolumn]{revtex4}
\usepackage{amssymb}
\usepackage{amsmath}
\usepackage{graphicx}
\usepackage{epsfig}

\begin{document}

\title{Quantum state swapping via qubit network with Hubbard interaction}
\author{S. Yang$^{1}$, Z. Song$^{1,a}$ and C. P. Sun$^{1,2,a,b}$ }
\affiliation{$^{1}$Department of Physics, Nankai University, Tianjin 300071, China}
\affiliation{$^{2}$ Institute of Theoretical Physics, Chinese Academy of Sciences,
Beijing, 100080, China}

\begin{abstract}
We study the quantum state transfer (QST) in a class of qubit network with
on-site interaction, which is described by the generalized Hubbard model
with engineered couplings. It is proved that the system of two electrons
with opposite spins in this quantum network of $N$ sites can be rigorously
reduced into $N$ one dimensional engineered single Bloch electron models
with central potential barrier. With this observation we find that such
system can perform a perfect QST, the quantum swapping between two distant
electrons with opposite spins. Numerical results show such QST and the
resonant-tunnelling for the optimal on-site interaction strengths.
\end{abstract}

\maketitle

\emph{Introduction. }For implementing quantum information processing based
on the scalable systems, the solid-state data bus is a necessary element to
coherently integrate two or more qubits and transfer the quantum information
among them \cite{Div}. Recently increasing investigations have explored the
possibilities to transfer quantum states through a class of solid-state data
buses, the artificial spin chain with engineered nearest neighbor (NN)
couplings \cite{Bose1,Bose2,Ekert1,Ekert2,ST,LY1,SZ,LY2}. Some novel
physical mechanisms have been discovered behind the protocols of quantum
state transfer (QST) based on the spin chain systems. For example, it is
discovered that the gap structure of spectrum of the strongly correlated
systems is responsible for the role of data bus \cite{LY1}; and the spectrum
marching parity symmetry is a sufficient condition for perfect QST \cite{ST,
LY2}.

We also notice that most of the explorations for QST are carried out only
for the non-interacting systems or single-particle quantum states. And it
seems that the on-site Coulomb interactions may destroy the quantum
coherence of transferred state. In this letter, we will study the influences
of on-site interactions on the dynamic process of QST by making use of the
generalized Hubbard model with engineered NN couplings as same as that in
the artificial Bloch electron model in ref. \cite{Ekert1}.

To give prominence to our central context we only consider the simplest
interacting system with only two electrons of opposite spin involved.\ The
main result we achieved is the discovery of the novel model reduction that
the $N$-site two-electron engineered Hubbard model can be decomposed into $N$
single-particle engineered models on $l$-sites chain ($l=1,3,...,2N-1$), but
with an additional central potential barrier (CPB). This discovery
enlightens us to conjecture the possibility of implementing the perfect
quantum information swapping since the reduced models still keep the mirror
symmetry. The detailed numerical simulations demonstrate there indeed exists
such perfect quantum state swapping even certain on-site repulsion $U$ is
considered.

%%%%%%%%%%%%%%%%%%%%%%%%%%%%%%%%%%%%%%%%%%%%%%%%%%%%%%%%%%%%%%%%%%%%%%%%%%%%%%%%%%%%%%%%%%%%%%%%%%%%%%%%%
\begin{figure}[tbp]
\includegraphics[bb=45 370 550 700, width=4 cm,clip]{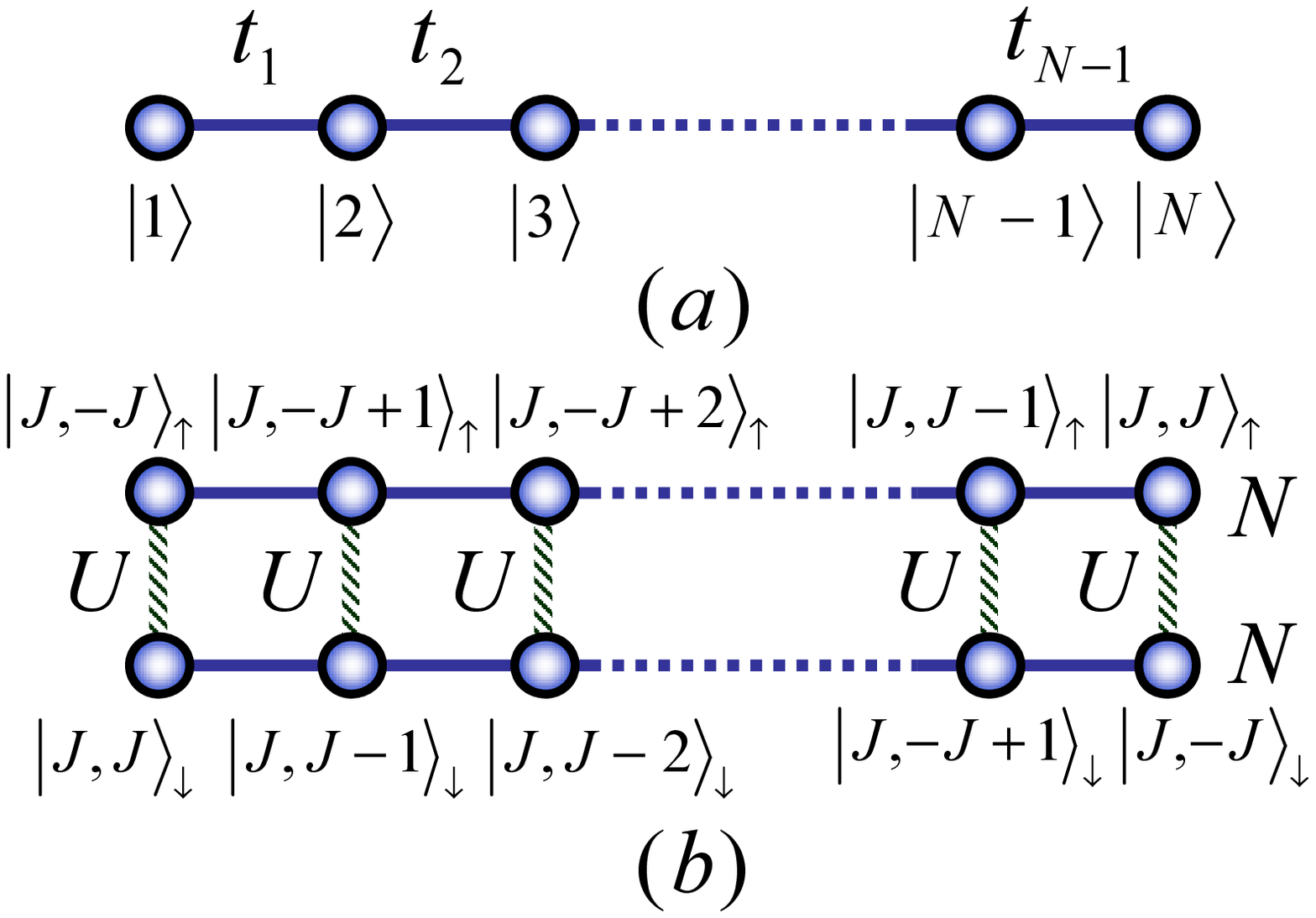} %
\includegraphics[bb=45 370 550 700, width=4 cm,clip]{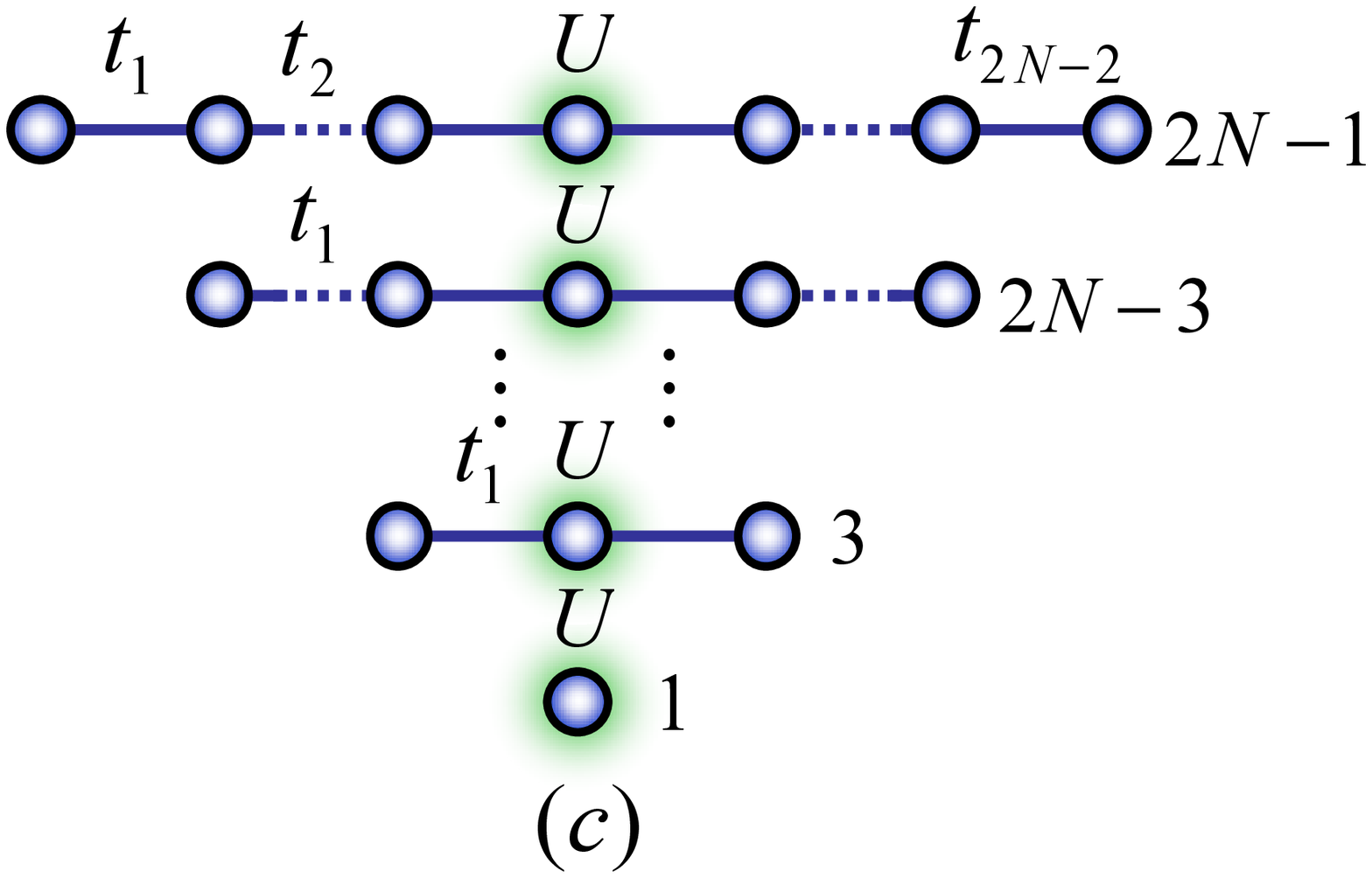}
\caption{\textit{(color on line) (a) The engineered Hubbard model of $N$
sites with two electrons. (b) The two-leg ladder of spinless Bloch
electrons, which is equivalent to the above Hubbard model. The rung
represents the on-site interactions. We label the sites in the two legs with
a standard angular momentum basis ${\left\vert JM\right\rangle }$ in two
opposite orders;(c) The above equivalent two-leg network can be further
reduced according to the product representations $SO(3)\otimes SO(3)$, as
the direct sum of $N$ central potential barrier model. }}
\end{figure}
%%%%%%%%%%%%%%%%%%%%%%%%%%%%%%%%%%%%%%%%%%%%%%%%%%%%%%%%%%%%%%%%%%%%%%%%%%%%%%%%%%%%%%%%%%%%%%%%%%%%%%%%%%

\emph{Engineered Hubbard model and its reduction.} Our model for quantum
state swapping is the generalizations of the engineered spin model in ref.
\cite{Ekert1} (see the Fig. 1a) by adding the on-site Coulomb interactions.
It can also be regarded as an engineered Hubbard model \cite{Hubbard} with
artificial hopping. The model Hamiltonian reads
\begin{equation}
H=-\sum_{j,\sigma }(t_{j}c_{j,\sigma }^{\dagger }c_{j+1,\sigma
}+h.c.)+U\sum_{j}n_{j\uparrow }n_{j\downarrow },
\end{equation}%
where $c_{j,\sigma }^{\dagger }$ is the creation operator of
electron at site $j$ with spin $\sigma =\uparrow ,\downarrow $ and
$U$ is the on-site repulsion. The hopping integral is engineered
as $\ t_{j}=\sqrt{j(N-j)}$. It has been widely studied in
connection with correlation effects in narrow-band solids and the
concept of entanglement \cite{Polao}. If
the on-site interaction is absent, i.e., $U=0$, it has been shown in \cite%
{Ekert1,Ekert2,LY2} that an arbitrary many-particle state can be transferred
to its mirror counterpart perfectly after the time $\tau =\pi /2.$ It is due
to the fact that the energy-level structure and the parity of the
corresponding eigenstate satisfy the spectrum-symmetry matching condition
(SSMC) introduced in \cite{LY2}: Let $\phi _{n}$ be the common
eigen-function of $H$ and mirror operator $R$ with the eigen-values $%
\varepsilon _{n}$ and $p_{n}$ respectively. It is easy to find that any
state $\psi $ at time $\tau $ can evolve into its symmetrical counterpart $%
R\psi $ if the eigenvalues $\varepsilon _{n}$ and $p_{n}$ match each other
and satisfy the SSMC $\exp (-i\varepsilon _{n}\tau )=p_{n}$.

It is easy to imagine that the energy levels should be shifted by ceratin
deviations from the original spectra when the on-site repulsion is switched
on. Nevertheless there still exists the possibility that the new set of
shifted energy levels satisfy the SSMC for an appropriate $U$ since the
certain symmetry remains as will be shown in the following discussions. We
first illustrate our analysis along this direction schematically in the
Suppose that there are only two electrons with opposite spin in the
engineered Hubbard model (see Fig. 1a). The on-site interaction occurs only
when the two electrons occupy a same site. Alternatively, the Hubbard chain
is equivalent to the spinless Bloch electron network with two legs and $N$
rungs (see Fig. 1b). The on-site interaction is denoted by the rungs. Each
site in the leg corresponds to a single electron Bloch state, $\left\vert
j\right\rangle _{\uparrow }=c_{j\uparrow }^{\dag }\left\vert 0\right\rangle $
or $\left\vert j\right\rangle _{\downarrow }=c_{j\downarrow }^{\dag
}\left\vert 0\right\rangle $, $j=1,2,\cdots ,N$.

According to the ref. \cite{Ekert1}, we can associate these states to the
angular momentum states%
\begin{equation}
\left\vert J,M\right\rangle _{\uparrow }=\left\vert J+M+1\right\rangle
_{\uparrow },\left\vert J,M\right\rangle _{\downarrow }=\left\vert
J-M+1\right\rangle _{\downarrow },
\end{equation}%
where a given $J=(N-1)/2,$ $M=J,J-1,\cdots ,-J+1,-J$. Then it is easy to
check that for the engineered couplings, the lowing operator of angular
momentum can be realized in terms of the fermion operators as%
\begin{eqnarray}
J_{-}^{(\uparrow )} &=&J_{x}^{(\uparrow )}-iJ_{y}^{(\uparrow
)}=\sum_{j}t_{j}c_{j,\uparrow }^{\dagger }c_{j+1,\uparrow } \\
J_{-}^{(\downarrow )} &=&J_{x}^{(\downarrow )}-iJ_{y}^{(\downarrow
)}=\sum_{j}t_{j}c_{j+1,\downarrow }^{\dagger }c_{j,\downarrow }
\end{eqnarray}%
which generates the group $SO(3)$ together with $J_{+}^{(\sigma )}$ $%
=(J_{-}^{(\sigma )})^{\dagger }$ and $J_{z}^{(\sigma )}=\sum_{j\sigma
}jc_{j,\sigma }^{\dagger }c_{j,\sigma }$ where $\sigma =\uparrow ,\downarrow
$. Then we can rewrite the Hamiltonian as%
\begin{eqnarray}
H &=&2J_{x}^{(\uparrow )}+2J_{x}^{(\downarrow )}+V;  \notag \\
V &=&U\sum_{M}\left\vert J,M;J,-M\right\rangle \left\langle
J,M;J,-M\right\vert ,
\end{eqnarray}%
where the two-particle associated state $\left\vert J,M;J,-M\right\rangle $
is defined by $\left\vert J,M;J,M^{\prime }\right\rangle $ $=$ $\left\vert
J,M\right\rangle _{\uparrow }\otimes \left\vert J,M^{\prime }\right\rangle
_{\downarrow }$.

The intrinsic dynamic symmetry of the above generalized model is described
as $SO(3)\otimes SO(3)$. Thus the addition theorem for two angular momenta
\cite{Beid} can be employed to reduce the representation of this generalized
Hubbard model according to the decomposition of the product representation $%
D^{[J]}\otimes D^{[J]}=$ $\sum_{L=0}^{2J}\oplus D^{[L]}$, where $D^{[J]}$ is
an irreducible representation of $SO(3)$. The key point in our treatment is
to expressed the on-site interaction term $V$ as the sum of irreducible
tensor operators. To this end, we use the Clebsch-Gordan coefficients $%
C_{J,M_{1};J,M_{2}}^{LM}=$ $\left\langle (JJ)L,JM\right. \left\vert
J,M_{1};J,M_{2}\right\rangle $ to write the eigenvector of the total angular
momentum $\left\vert L,M\right\rangle \equiv $ $\left\vert
(JJ)L,JM\right\rangle =$ $\sum_{M_{1}+M_{2}=M}C_{J,M_{1};J,M_{2}}^{LM}\left%
\vert J,M_{1};J,M_{2}\right\rangle $. From the corresponding inverse
transformation, the interaction term can be decomposed as%
\begin{equation}
V=U\sum_{LL^{\prime }}\sum_{M}C_{J,M;J,-M}^{L0}C_{J,M;J,-M}^{L^{\prime
}0}\left\vert L,0\right\rangle \left\langle L^{\prime },0\right\vert .
\end{equation}%
Due to the orthogonal relation of Clebsch-Gordan coefficients%
\begin{equation}
\sum_{m}C_{J,m;J,M-m}^{LM}C_{J,m;J,M-m}^{L^{\prime }M}=\delta _{L,L^{\prime
}},
\end{equation}%
the on-site interaction can be reduced as the sum of the irreducible
tensors, i.e., $V=\sum_{L}W^{[L]}=$ $U\sum_{L}\left\vert
(JJ)L,0\right\rangle \left\langle (JJ)L,0\right\vert $.

Therefore, we have proved that the engineered Hubbard Hamiltonian can be
written as the direct sum of $N$ irreducible sub-Hamiltonians $%
H^{(L)}=H_{0}^{(L)}+W^{[L]}$ $=2J_{x}+W^{[L]}.$ The model described by each $%
H^{(L)}$ can be inversely mapped into a new Bloch electron model with a CPB,
whose Hamiltonian is
\begin{equation}
H^{(L)}=\sum_{j=-L}^{L-1}(t_{j}a_{j}^{\dag }a_{j+1}+h.c.)+Ua_{0}^{\dag
}a_{0},  \label{CPB}
\end{equation}%
where $a_{j}^{\dag }$ is the creation operator of new fermion and $t_{j}=$ $%
\sqrt{(L+j+1)(L-j)}$. As illustrated in Fig. 1c, the on-site interacting
qubit network is reduced into a direct sum of $N$\ Bloch electron models
with CPB.

\emph{Spectrum-symmetry matching for nonzero }$\emph{U}$\emph{.}$\ $To see
whether the engineered Hubbard model can serves as a quantum data bus to
coherently transfer quantum information, we need to study the influences of
the on-site interaction on the SSMC.

Let us first recall the on-site interaction free case,\ where the matrix
representation of the Hamiltonian (1) in single-particle subspace is
equivalent to that a high spin with angular momentum $J=(N-1)/2$ precessing
in transverse magnetic field. For two-electron case, the above analysis
shows that the original Hamiltonian can be reduced into $N$ single-particle
engineered models on $l$-site chain ($l=1,3,...,2N-1$) without CPB.
Obviously, all the eigenstates also meet the SSMC that guarantees a perfect
QST. It is crucial for our analysis that the mirror symmetry is not broken
even in presence of the on-site interaction, then the effect of nonzero $U$
on the deviation of the energy levels determines the fidelity of QST via
such system.

Now we consider the Hamiltonian (\ref{CPB}) with nonzero $U$. Since $%
[H^{(L)},R]=0$, the eigenstates can be classified into two sets with
different parities. When $U$ is switched on, a set of levels reminds
unchanged while other set of levels deviates by a nonzero values $\Delta
_{M}=|E_{M}(U)-E_{M}(0)|$, (see Fig. 2). To prove this generally, we
calculate the action of $W^{[L]}$ on the eigenvectors
\begin{equation}
\left\vert L,M(\frac{\pi }{2})\right\rangle =e^{i\frac{\pi }{2}%
J_{y}}\left\vert L,M\right\rangle =\sum_{M^{\prime }}d_{M^{\prime }M}^{L}(%
\frac{\pi }{2})\left\vert L,M^{\prime }\right\rangle
\end{equation}%
of the reduced Hamiltonian $H_{0}^{(L)}=2J_{x}$. Because $d_{M^{\prime
}M}^{L}(\pi /2)=$ $(-1)^{L-M}d_{-M^{\prime }M}^{L}(\pi /2)$ we have $%
d_{0M}^{L}(\pi /2)=0$ for odd $L-M$, and then
\begin{equation}
W^{[L]}\left\vert L,M(\frac{\pi }{2})\right\rangle =Ud_{0M}^{L}(\frac{\pi }{2%
})\left\vert L,0\right\rangle =0.
\end{equation}%
This means that states $\left\vert L,M(\pi /2)\right\rangle $ ($L-M$ is odd)
are also the eigenstates of $H^{(L)}$. Indeed these corresponding levels are
free of the on-site interaction. %
%%%%%%%%%%%%%%%%%%%%%%%%%%%%%%%%%%%%%%%%%%%%%%%%%%%%%%%%%%%%%%%%%%%%%%%%%%%%%%%%%%%%%%%%%%%%%%%%%%%%%%%%%
\begin{figure}[tbp]
\includegraphics[bb=90 307 457 772, width=6 cm, clip]{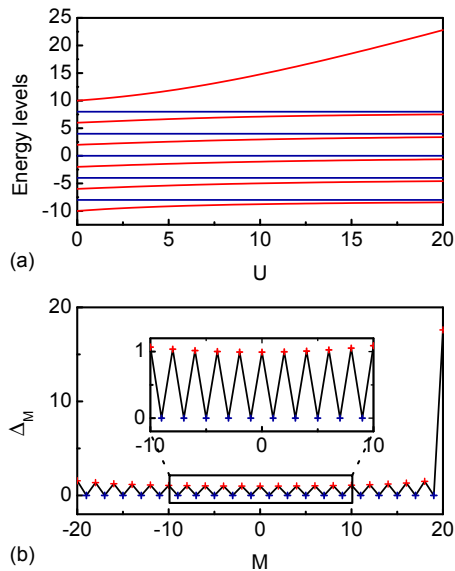}
\caption{\textit{(color on line) (a) Numerical simulation of the energy
levels effected by $U$ for $H^{(L)}$ with $L=5$, $U=0\sim 20$. (b) The level
shift $\Delta _{M}$ of $H^{(L)}$ for $L=20$ system. It shows that for small $%
M$, $\Delta _{M}$ is approximately uniform. Notice that for the optimal $%
U=40.5$, the level shifts for small $M$ are approximately equal to the half
of the level difference with $U=0$.}}
\end{figure}
%%%%%%%%%%%%%%%%%%%%%%%%%%%%%%%%%%%%%%%%%%%%%%%%%%%%%%%%%%%%%%%%%%%%%%%%%%%%%%%%%%%%%%%%%%%%%%%%%%%%%%%%%%

Another set of eigenvalues should be shifted by the on-site interaction.
Imagine that if the energy deviation$\ \Delta _{M}$ is not so sensitive to $%
M $, the shifts of the levels are approximately the same. There may exist an
appropriate $U$ to ensure that the final levels satisfy the SSMC with
another greatest common divisor. It will result in the perfect QST in the
invariant subspace $V^{[L]}:$ $\{\left\vert L,M\right\rangle ,$ $M=L,$ $%
L-1,\cdots ,-L\}$.

In order to verify our conjecture, numerical simulation is employed for
small size systems. Exact diagonalization results for $H^{(L)}$ on $L$
lattice are plotted in Fig. 2. In Fig. 2a, the energy levels as functions of
$U$ for $L=5$ system shows that the spectrum of $H^{(L)}$ consists of two
sets of energy levels, one is independent of $U$, while the other is shifted
by the repulsion. This conclusion is in agreement with the above analysis
for arbitrary $L$. Numerical calculation for $\Delta _{M}$ of $H^{(L)}$ with
$U=40.5$ on $21$-site lattice is plotted in Fig. 3b. It shows that for small
$M$, $\Delta _{M}$ is approximately uniform. On the other hand, numerical
calculations also indicate that in the invariant subspace $V^{[L]}$, the
components of the state $\left\vert L,-L\right\rangle $ on the basis $%
V^{[L]} $ for small $M$ are dominant, i.e, the effective levels of such
state should be shifted uniformly. In other words, for such kind of initial
state $\left\vert L,-L\right\rangle $, when $U$ takes an appropriate value,
the effective levels can satisfy the SSMC approximately. Thus state $%
\left\vert L,-L\right\rangle $\ can be transferred into $\left\vert
L,L\right\rangle $ near perfectly.

\emph{Near-perfect swap of two electrons.} Now we consider the
spin state swapping of two electrons located on the two ends of
the one-dimensional lattice. For the initial state
\begin{equation}
\left\vert \psi (0)\right\rangle =C_{1,\uparrow }^{\dagger }C_{N,\downarrow
}^{\dagger }\left\vert 0\right\rangle =\left\vert J,-J;J,-J\right\rangle ,
\end{equation}%
the quantum state swapping is a mapping from $\left\vert \psi
_{J}(0)\right\rangle $ to
\begin{equation}
\left\vert \psi (t)\right\rangle =C_{N,\uparrow }^{\dagger }C_{1,\downarrow
}^{\dagger }\left\vert 0\right\rangle =-\left\vert J,J;J,J\right\rangle
\end{equation}%
can be approximately realized in a dynamic process since the SSMC can be
satisfied for an appropriate $U$ as discussed above.

In order to confirm the above prediction, the numerical simulation is
performed for the swapping fidelity
\begin{equation}
F(U,t)=\left\vert \left\langle J,J;J,J\right\vert e^{-iHt}\left\vert
J,-J;J,-J\right\rangle \right\vert ^{2}
\end{equation}%
where $H$ is the Hamiltonian (1) for the engineered Hubbard model. During
the time range $t\in $ $[0,10]$, the maxima of the fidelity, $F_{\max }(U)=$
$\max \{F(U,t),$ $t\leqslant 10\}$ are plotted in Fig. 3a as the functions
of the on-site interaction strength $U$ for the system of $N=4,5$ and $6$
sites. It shows that there indeed exist some $U$ to get very high $F_{\max
}(U)$, which seems like that the propagation of electrons is
scattering-free. One can see for the same $N$ that $F_{\max }$ have several
regular peaks, and for each peak, the revival times $T_{r}$\ are different.
It is interesting to find that, for the first regular peak, the revival time
$T_{r1}\simeq \pi $, the second peak, $T_{r2}\simeq 1.5\pi $, ..., and the $%
N $th peak, $T_{rn}\simeq 0.5\pi (n+1)$. The corresponding optimized $U$ and
revival time $T_{r}$ for each peak are listed in Table 1.

\begin{center}
\begin{tabular}{cccc}
\hline\hline
Peaks & $\ \ U$ \ \  & $\ \ F_{\max }$ \ \  & $\ \ \ \ \ \ \ \ \ T_{r}$ \ \
\ \ \ \ \  \\ \hline
$1$ & $6.6$ & $0.9847$ & $3.14\simeq \pi $ \\
$2$ & $11.6$ & $0.9768$ & $4.71\simeq 3\pi /2$ \\
$3$ & $16.2$ & $0.9724$ & $6.28\simeq 2\pi $ \\
$4$ & $20.6$ & $0.9698$ & $7.85\simeq 5\pi /3$ \\
$5$ & $25.0$ & $0.9683$ & $9.42\simeq 3\pi $ \\ \hline\hline
\end{tabular}
\end{center}

\vspace{0.2cm}

\begin{center}
Table 1
\end{center}

\textit{Table 1. The maxima of fidelity $F_{\max }$, the corresponding
optimized $U$ and revival time $T_{r}$ obtained by numerical simulations for
the engineered Hubbard model on $4$-site system.} \newline
\newline
On the other hand, for different $N=2\sim 10$, more detailed data, such as
the optimized $U$, $F_{\max }$ and the corresponding revival time $T_{r}$
for the first and second peaks are also obtained numerically to\ reveal the
hidden relationship between them. In Table 2, these numerical results are
listed with some obvious characters for the cases of $N=4 \sim 10$.

\begin{center}
\begin{tabular}{ccccccc}
\hline\hline
& \multicolumn{3}{c}{1st Peaks} & \multicolumn{3}{c}{\ 2nd Peaks} \\ \hline
$\ \ \ \ N$ \ \ \  & $\ \ U$ \ \  & $\ F_{\max }$ \  & $\ T_{r}$ \  & $\ \ \
\ U$ \ \  & $\ F_{\max }$ \  & $\ T_{r}$ \ \  \\ \hline
$2$ & $2.3$ & $0.9999$ & $2.72$ & $\ 3.6$ & $0.9999$ & $3.50$ \\
$3$ & $4.9$ & $0.9926$ & $3.18$ & $\ 8.0$ & $0.9929$ & $4.74$ \\
$4$ & $6.6$ & $0.9847$ & $3.14$ & $\ 11.6$ & $0.9768$ & $4.71$ \\
$5$ & $8.6$ & $0.9873$ & $3.14$ & $\ 15.0$ & $0.9802$ & $4.71$ \\
$6$ & $10.6$ & $0.9906$ & $3.14$ & $\ 18.4$ & $0.9856$ & $4.71$ \\
$7$ & $12.6$ & $0.9931$ & $3.14$ & $\ 21.8$ & $0.9894$ & $4.71$ \\
$8$ & $14.5$ & $0.9948$ & $3.14$ & $\ 25.2$ & $0.9920$ & $4.71$ \\
$9$ & $16.5$ & $0.9960$ & $3.14$ & $\ 28.7$ & $0.9938$ & $4.71$ \\
$10$ & $18.5$ & $0.9968$ & $3.14$ & $\ 32.1$ & $0.9950$ & $4.71$ \\
\hline\hline
\end{tabular}
\end{center}

\vspace{0.2cm}

\begin{center}
Table 2
\end{center}

\textit{Table 2. The maximal fidelities and the corresponding revival time $%
T_{r}$ of the first two peaks obtained by numerical simulations for the
systems with $N$ and $U$.} \newline
\newline

As for \ the quantum swapping scheme we study here, the following prominent
characters can be found from the above analysis. Firstly, the optimized $U$
is linearly proportional to $N$ with slopes approximately $2$ and $3.4$ for
the first and second peak respectively. Secondly, for a given optimized $U$,
the larger $N$ is, the higher the\ maximum of the fidelity $F_{\max }$
becomes. Consequently, the QST will be better. Thirdly, as the statements
above, the revival times are $\pi $ and $3\pi /2$ for the first and second
peak, approximately. Our further numerical results shows that the above
experiential law can hold for larger $N$. For different peaks, the linear
relations between optimized $U$ and the sizes $N$ are also shown in Fig. 3b.
%%%%%%%%%%%%%%%%%%%%%%%%%%%%%%%%%%%%%%%%%%%%%%%%%%%%%%%%%%%%%%%%%%%%%%%%%%%%%%%%%%%%%%%%%%%%%%%%%%%%%%%%%
\begin{figure}[tbp]
\includegraphics[bb=20 310 480 780, width=4 cm, clip]{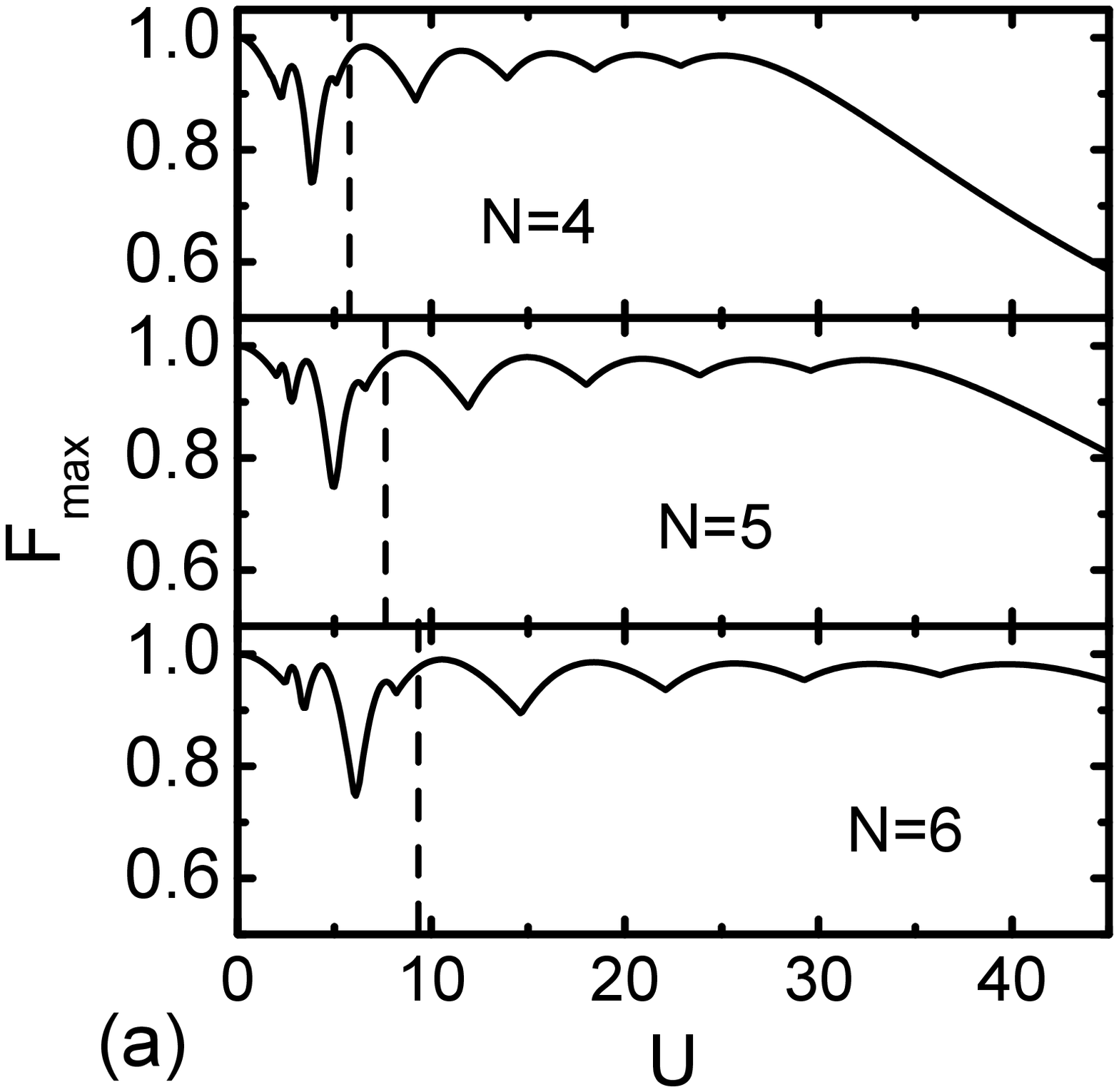} %
\includegraphics[bb=40 310 500 780, width=4 cm, clip]{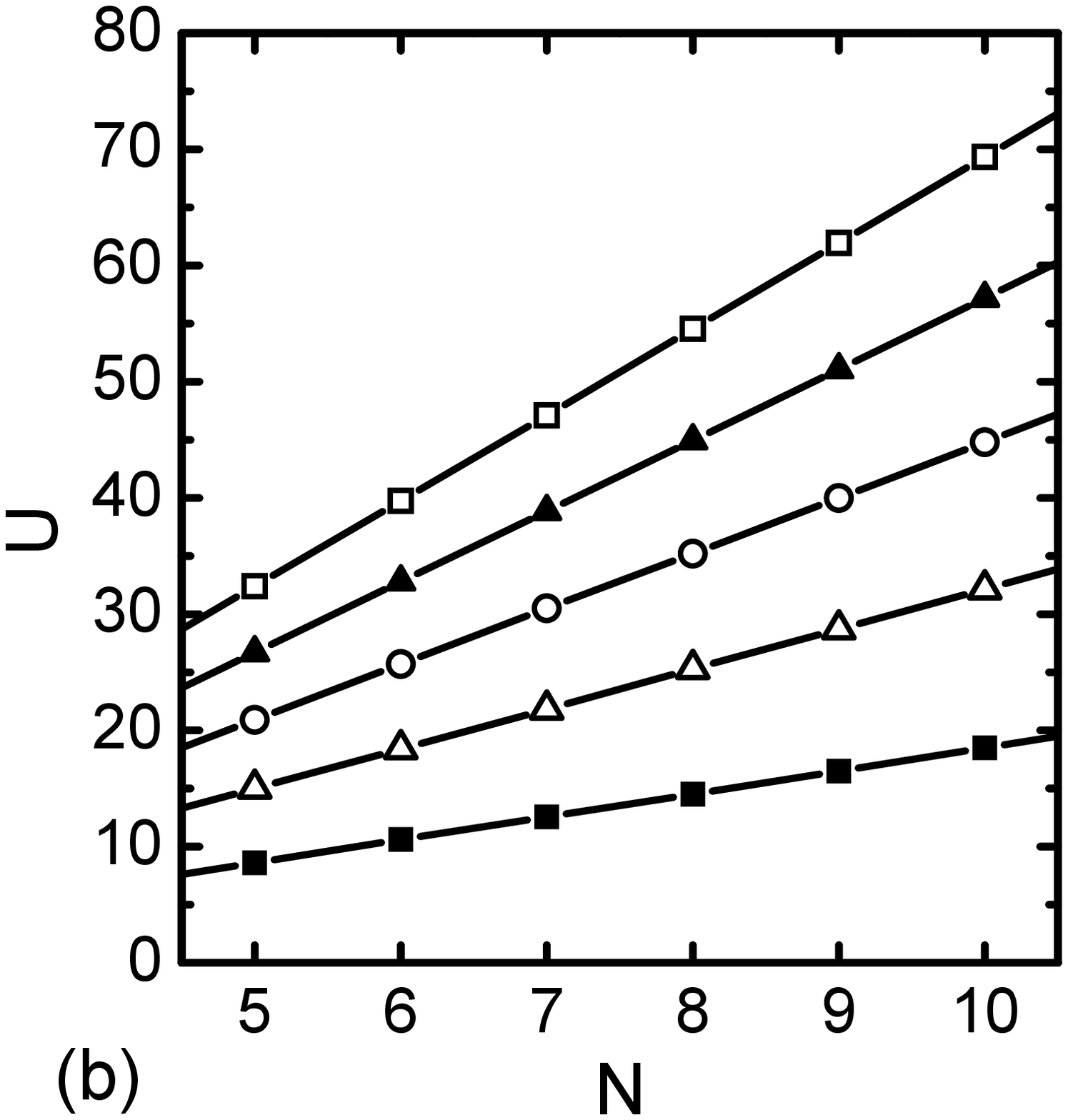}
\caption{\textit{(a) Numerical simulation of the maximal $F_{\max }$ as the
function of the interaction strength $U$ for $N=4$ sites system. (b) The
optimal $U$ as the functions of the sizes $N$ obtained by the numerical
simulations. The corresponding revival times, $T_{r}=\protect\pi $ (solid
square), $\protect\pi /2$ (triangle), $2\protect\pi $ (circle), $5\protect%
\pi /2$ (solid triangle), and $3\protect\pi $ (square). It indicates that
the optimal interaction strengths $U$ are directly proportional to $N$
approximately.}}
\end{figure}
%%%%%%%%%%%%%%%%%%%%%%%%%%%%%%%%%%%%%%%%%%%%%%%%%%%%%%%%%%%%%%%%%%%%%%%%%%%%%%%%%%%%%%%%%%%%%%%%%%%%%%%%%%

From the Tables we set above, it is obvious that the revival times obey the
experience formula $T_{rn}=0.5\pi (n+1)$ approximately, where $n=1,2,...,$
is the order of the peaks of $F_{\max }$. This formula can be understood
based on the above analysis. If the effective levels are shifted by $U$
uniformly from $\Delta _{M}=0$ to $1$, the possible greatest common divisors
meet the SSMC are $2/(n+1)$ ($n=1,2,...,$). The corresponding level shift is
$2[1-1/(n+1)]$, which results in the revival period $T_{rn}$. A question to
be asked is about the relationship between the level shifts and the
repulsion $U$. In Fig. 3b, the optimized $U$ as the functions of the sizes $%
N $ are plotted. Interestingly, they are simply linear functions in the
range we concerned.

\emph{Summary. }In summary, we study the QST in the engineered Hubbard model
with on-site interaction analytically and numerically. It is proved that the
system of two electrons with opposite spins in this $N$ sites quantum
network can be reduced into $N$ one dimensional engineered single Bloch
electron models with CPB rigorously. Analytic calculation and numerical
results both show that the engineered Hubbard model can perform perfect
quantum swapping between two distant electrons with opposite spins even when
the certain Coulomb interaction exists. The angular momentum reduction
method could be expected to work for the QST in an engineered quantum spin
models with more electrons.

This work is supported by the NSFC with grant Nos. 90203018, 10474104 and
60433050. It is also funded by the National Fundamental Research Program of
China with Nos. 2001CB309310 and 2005CB724508.

\end{document}